\documentstyle[aps,preprint]{revtex}
\begin{document}
\draft
\title{Effect of disorder on the magnetic and transport properties of La$_{1-x}$Sr$%
_x$MnO$_3$}
\author{R. Allub\thanks{%
Member of the Carrera del Investigador Cient\'{\i}fico del Consejo Nacional
de Investigaciones Cient\'{\i}ficas y t\'{e}cnicas (CONICET).} and B. Alascio%
$^{*}$}
\address{Centro At\'{o}mico Bariloche, (8400) S. C. de Bariloche, Argentina.}

\maketitle

\begin{abstract}
We study a simplified model of the electronic structure of compounds of the
type of La$_{1-x}$Sr$_x$MnO$_3$. The model represents each Mn$^{4+}$ ion by
a spin S=1/2, on which an electron can be added to produce Mn$^{3+}$. We
include two strong intratomic interactions in the Hamiltonian: exchange ($J$%
) and Coulomb ($U$). Finally, to represent the effect of Sr substitution by
La in a simple way, we include a distribution of diagonal energies at the Mn
sites. Then we use Green function techniques to calculate a mobility edge
and the average density of states. We find that according to the amount of
disorder and to the concentration of electrons in the system, the Fermi
level can cross the mobility edge to produce a metal to insulator transition
as the magnetization decreases (increase of temperature). If the disorder is
large, the system remains insulating for all concentrations. Concentrations
near zero or one favor the insulating state while intermediate values of
concentration favor the metallic state.

Keywords: A. disordered systems A. magnetically ordered materials
          D. electronic transport

\end{abstract}

\pagebreak

The recent discovery of extremely large magneto resistance in La$_{1-x}$A$_x$%
MnO$_3$ (where A stands for Ca, Sr, and Ba) and other related oxides \cite{1}
has renewed interest in these types of compounds. The magneto resistance
values found at large fields seem to be connected with a metal-insulator
transition found in some of the compounds at temperatures equal or lower
than the ferromagnetic to paramagnetic transition. The $T$ vs. $x$ phase
diagram of these compounds is quite similar to that of the high temperature
superconductors at low and intermediate concentrations of A, but with the
ferromagnetic phase substituting the superconducting one. At the end of the
composition range both LaMnO$_3$ and CaMnO$_3$ are antiferromagnetic
insulators, while in the intermediate composition range the compounds are
ferromagnetic metals. Much has been learned about these compounds and alloys
since the earlier studies of Jonker and Van Santen \cite{2}. The later
experimental results including single crystal measurements are listed in
Ref. 3. Zener \cite{4} advanced a mechanism to understand the properties of
the different compounds. Based on the coexistence of Mn$^{3+}$ (3d$^4$) and
Mn$^{4+}$ (3d$^3$) within the doped materials, he proposed that the
displacement of the fourth electron between them produces the metallic
conductivity and at the same time provides a mechanism for the ferromagnetic
alignment of the spins. Anderson and Hasegawa \cite{5} calculated the
interaction between two magnetic ions mediated by the ''double exchange
mechanism''. They point out the importance of the orbital degeneracy for
this mechanism as is the case in the Mn ions. de Gennes \cite{6} made a
study of the competition between superexchange and double exchange in the
dilute limit ($x<<1$) compounds and proposed the existence of canted or
spiral magnetic structures. Kubo and Ohata \cite{7} used a spin wave
approach to study the temperature dependence of the resistivity well below
the Curie Temperature and a mean field approach to the many body Hamiltonian
to obtain the electronic and magnetic excitations in the metallic phase.
Mazzaferro, Balseiro, and Alascio \cite{8} used a strong coupling approach
based on the similarity of these compounds to the intermediate valence Tm
compounds. They introduced in the model the possibility of non- equivalent
sites for the Mn ions. To our knowledge, this was the first time that the
possibility of a metal to insulator transition was mentioned in connection
with double exchange. Since magnetization and conductivity are intimately
connected it is evident that any appropriate description of the properties
of these compounds must include those factors that affect the mobility of
the carriers.

Furukawa used the infinite dimension Kondo lattice model with classical
spins to describe several properties of the compounds. Within the model, he
was able to fit the resistance versus magnetization curves of Tokura et. al.
\cite{9} showing that double exchange is essential to the theory of doped
LaMnO$_{3}$. Millis et. al. \cite{10} analyze a 'ferromagnetic Kondo
Hamiltonian' using different approximate methods to conclude that double
exchange alone is not sufficient to describe the properties of La$_{1-x}$Sr$%
_x$MnO$_3$.

In this paper we extend the study of Ref. 8 including a continuous
distribution of inequivalent sites for the Mn ions. We find that as the
magnetization increases from zero to saturation the system may undergo a
transition from a state where the Fermi level falls below the localization
edge to one where it falls above the edge thus substantially changing the
transport properties.

We consider a model were one electron can hop from the 3d$^4$ configuration
of Mn$^{3+}$ to the 3d$^3$ configuration of a Mn$^{4+}$ nearest neighbor. To
consider the Zener double exchange mechanism, we include an exchange energy
between the fourth electron and a core formed by the three localized
electrons in the lower d orbitals at each site. To avoid double occupation
by itinerant electrons, we take a large Coulomb interaction between them.
Finally to simulate the effect of disorder produced by the substitution of
La by Sr or other divalent ions and other defects, we introduce a
distribution of diagonal energies for different sites. We use a very
simplified model Hamiltonian that represents each Mn$^{4+}$ ion at site $i$
by a spin $S_i$, on which one electron can be added to produce Mn$^{3+}$
(for simplicity in what follows we take $S_i=\frac 12$). When an electron is
added in the d-shell of site $i$, an exchange coupling $J$ is included to
favor parallel alignment of the added electron to the already existing spin
\cite{11}. Also to avoid the possibility of Mn$^{2+}$ we include a strong
Coulomb repulsion $U$. The Hamiltonian reads: \pagebreak
$$
H=\sum_{i,\sigma }\epsilon _ic_{i\sigma }^{\dagger }c_{i\sigma
}-t\sum_{<i,j>,\sigma }c_{i\sigma }^{\dagger }c_{j\sigma }
$$
\begin{equation}
\hspace{10mm}+U\sum_ic_{i\uparrow }^{\dagger }c_{i\uparrow }c_{i\downarrow
}^{\dagger }c_{i\downarrow }-J\sum_i\vec {S_i}.\vec {\sigma _i}\,,
\end{equation}
where $c_{i\sigma }^{\dagger }$, $c_{i\sigma }$ creates and destroys an
itinerant electron with spin $\sigma $ at site $i$, respectively. $\vec {S_i}
$ and $\vec {\sigma _i}$ are the Pauli matrices for spin $\frac 12$ at site $%
i$ for localized and itinerant electrons respectively. $\epsilon _i$ is the
on-site energy, $t$ the hopping parameter between nearest neighbors, $U$ the
on-site Coulomb repulsion between two itinerant electrons, and $J$ is the
ferromagnetic ($J>0$) coupling between the localized and itinerant
electrons. As is well known, this coupling does not give rise to divergence
in the impurity scattering as in the Kondo case nor to heavy electron
dynamics as in the Kondo compounds. Without losing essential physics we
simplify further by taking only the $z$ component of the exchange
interaction. Thus the states of the system are characterized by itinerant
electrons moving on a frozen distribution of localized up or down spins. To
obtain site Green functions and thus local density of states for this
problem, we start from Hamiltonian (1) for $t=0$ and we calculate the local
Green function. For spin up we obtain:
\begin{equation}
G_{i\uparrow \alpha }^0=<<c_{i\uparrow };c_{i\uparrow }^{\dagger }>>=\frac{%
[\omega -E_{i\alpha }-U(1-\bar n_{i\downarrow })]}{(\omega -E_{i\alpha
})(\omega -E_{i\alpha }-U)}\,,
\end{equation}
where $\bar n_{i\downarrow }=<c_{i\downarrow }^{\dagger }c_{i\downarrow }>$,
$\alpha =+$ ($\alpha =-$) for up (down) localized spin, and $E_{i\alpha
}=\epsilon _i-\alpha J$. In order to obtain an approximate solution to the
problem for ($t\neq 0$) at the start we ignore the site dependence of the
diagonal energies: i.e. we set $\epsilon _i=\epsilon $ and we are left with
a binary alloy problem. Using the Renormalized Perturbation Expansion (RPE)
\cite{12} and Eq. (2) for $U\to \infty $ we obtain the corresponding local
Green functions. Spin up gives
\begin{equation}
G_{i\uparrow \alpha }=\frac{(1-\bar n_{i\downarrow })}{[\omega -E_{i\alpha
}-(1-\bar n_{i\downarrow })\Delta _{i\uparrow }]}\,,
\end{equation}
where $\Delta _{i\uparrow }$ is the corresponding self-energy given by
\begin{equation}
\Delta _{i\uparrow }=t^2\sum_{\delta \alpha }G_{i+\delta \uparrow \alpha
}^{\prime }\,,
\end{equation}
the summation over $\delta $ runs over the nearest neighbors to site $i$ and
the $G^{\prime }$ are the local propagators avoiding $i$. Their
self-energies $\Delta _{i+\delta \uparrow }^{\prime }$ are given by
\begin{equation}
\Delta _{i+\delta \uparrow }^{\prime }=t^2\sum_{\delta ^{\prime }\alpha
}G_{i+\delta +\delta ^{\prime }\uparrow \alpha }^{\prime \prime }\,,
\end{equation}
and so on. Thus, for a system involving an infinite number of sites the RPE
iterated procedure as indicated above gives an infinite number of steps. In
order to get an approximate solution we average over spin configurations the
second term in Eqs.(4) and (5) which gives
$$
\Delta _{\uparrow }=(K+1)t^2[\frac{(1-\bar n_{\downarrow }^{+})\nu _{+}}{%
[\omega -E_{+}-(1-\bar n_{\downarrow }^{+})\Delta _{\uparrow }^{\prime }]}
$$
\begin{equation}
\hspace{20mm}+\frac{(1-\bar n_{\downarrow }^{-})\nu _{-}}{[\omega -E_{-}-(1-%
\bar n_{\downarrow }^{-})\Delta _{\uparrow }^{\prime }]}]\,,
\end{equation}
where $E_\alpha =(\epsilon -\alpha J)$, $\bar n_{\downarrow }^{+}$ ($\bar n%
_{\downarrow }^{-}$) is the average number of itinerant electrons with spin
down at localized spin up (down) sites, $(K+1)$ is the number of nearest
neighbors (six for the simple cubic $Mn$ lattice), $\nu _{+}$ ($\nu _{-}$)
is the probability that a site has up (down) localized spin (notice that $%
(\nu _{+}+\nu _{-})=1$), and
$$
\Delta _{\uparrow }^{\prime }=Kt^2[\frac{(1-\bar n_{\downarrow }^{+})\nu _{+}%
}{[\omega -E_{+}-(1-\bar n_{\downarrow }^{+})\Delta _{\uparrow }^{\prime }]}
$$
\begin{equation}
\hspace{15mm}+\frac{(1-\bar n_{\downarrow }^{-})\nu _{-}}{[\omega -E_{-}-(1-%
\bar n_{\downarrow }^{-})\Delta _{\uparrow }^{\prime }]}]\,.
\end{equation}
Notice that within this approximation the iterative procedure closes in the
last Equation. This allows us to obtain average density of states for spin
up and down. At $J>\sqrt{K}t$ the density of states splits into two bands
centered at $E_\alpha $ with weights and widths that depend on the number of
sites with each spin. i.e. they depend on the magnetization of the system.
In Fig. 1 we show the densities of states for spin up and down for several
values of the magnetization ($m$), with $m=(\nu _{+}-\nu _{-})$. For $J>>%
\sqrt{K}t$, the term proportional to $\nu _{-}$ can be neglected in Eq.(6)
and (7). Similarly for $\Delta _{\downarrow }$ and $\Delta _{\downarrow
}^{\prime }$, consequently it is easy to see that $n_{\downarrow }^{+}\to 0$
and $n_{\uparrow }^{-}\to 0$. Eq.(7) reduces to
\begin{equation}
\Delta _{\uparrow }^{\prime }=\frac{(\omega -E)}2\pm \sqrt{\frac{(\omega
-E)^2}4-Kt^2\nu _{+}}\,,
\end{equation}
where $E=(\epsilon -J)$ and $\Delta _{\uparrow }$ results in this case
\begin{equation}
\Delta _{\uparrow }=\frac{(K+1)t^2\nu _{+}}{[\frac{(\omega -E)}2\mp \sqrt{%
\frac{(\omega -E)^2}4-Kt^2\nu _{+}}]}\,.
\end{equation}
Finally we obtain
\begin{equation}
G_{\uparrow +}=\frac{(K-1)(\omega -E)\mp (k+1)\sqrt{(\omega -E)^2-4Kt^2\nu
_{+}}}{2[(K+1)^2t^2\nu _{+}-(\omega -E)^2]}\,.
\end{equation}
Eq. (8) allow us to obtain the density of states per site as
\begin{equation}
\rho _{0\uparrow +}(\omega )=\frac{(K+1)\sqrt{4Kt^2\nu _{+}-(\omega -E)^2}}{%
2\pi |(K+1)^2t^2\nu _{+}-(\omega -E)^2|}\,.
\end{equation}
At this point, we introduce the site dependent diagonal energies. As is well
known, since Anderson's original paper \cite{13} , a distribution of
diagonal energies produces localization of the electronic states from the
edges of the bands to an energy within them which is called ''mobility
edge'' (ME). The precise position of the ME is difficult to calculate,
different localization criteria result in different values for it \cite{14}.
Our concern here is with the changes of the ME with magnetization, which do
not differ much among the different criteria. For this reason we report
results using Anderson's original criterium.

Since the resulting structure of localized and extended states in each band
can only be qualitatively obtained after having tried several distribution
of energies, we present the simplest mathematical model for the
distributions of energies: i.e. a Lorentzian \cite{15} distribution of width
$\Gamma $. The ensemble-averaged Green function allows us to write an
approximate density of states given by
\begin{equation}
\rho _{\uparrow +}(\epsilon )=\int_{-\infty }^{+\infty }\rho _{0\uparrow
+}(\epsilon ^{\prime })L(\epsilon -\epsilon ^{\prime })d\epsilon ^{\prime
}\,,
\end{equation}
where $L(x)$ is a Lorentz distribution given by
\begin{equation}
L(x)=\frac \Gamma {[\pi (x^2+\Gamma ^2)]}\,.
\end{equation}
In a similar manner we obtain $\rho _{\downarrow -}$.

In Fig. 2 we show the resulting densities of states and mobility edges for
different values of the magnetization. Inspection of Fig.2 shows that the
system can change from metallic at full magnetization (the Fermi level falls
above the mobility edge of the majority band) to localized at the
paramagnetic state (the Fermi level below the mobility edge of the same
band). Depending on the values of $\Gamma $ and $n$ , situations where the
system remains localized or metallic for all values of the magnetization can
also be found. This possibility of the system of changing character of
states at the Fermi level from extended to localized can be summarized in
the phase diagram depicted in Fig.3, where the localized or extended
character of the states at the Fermi level are shown as functions of the
electron concentration and magnetization for different values of $\Gamma $.

We do not derive here an expression for the Free energy of the system that
would allow us to calculate different thermodynamic properties, however we
can easily estimate the effect of disorder on the Curie temperatures: The
energy difference between the zero-temperature, fully polarized system and
the paramagnetic state can be calculated from the density of states and is a
function of the disorder parameter $\Gamma $ and the bandwidth. For $\Gamma
>>\sqrt{K}t$ the total density of states (up and down bands) is weakly
dependent on $m$ while in the opposite case the density of states changes
substantially as $m$ decreases from saturation to zero. The entropy change
between paramagnetic and ferromagnetic states, however, is practically
independent of the parameters in the Hamiltonian and is dominated by the
spin entropy and approximately equal to $k_B\,ln(3+n)$. In fig 4. we show
the electronic energy as a function of the $n$ for two values of the
disorder parameter. The Curie temperatures estimated using the above
argument range from 0K to about 800K at the top of the scale. One can see
that the Curie temperatures decrease with increasing $\Gamma $.

The absence of Mn$^{2+}$ and the value of the saturation magnetization
indicate that our assumptions $U,J>>t,\Gamma$ are valid in most of the Mn
oxides under consideration. The low energy properties then, depend only on $%
t $, $n$, and $\Gamma$. Of these, $\Gamma$ is the most difficult parameter
to estimate, as it should include ${\it {all}}$ effects of disorder. Among
these, substitution of La by divalent ions and O vacancies are the most
evident sources of disorder, but other types of defects also contribute to $%
\Gamma$. Polaronic effects derived from Jahn-Teller as well as breathing
phonons should be present in these materials. It is conceivable that they
could also be represented within our assumptions of a distribution of
diagonal energies.

We propose here a model that allows to understand the properties of the
different magnetoresistive compounds in terms of a disorder parameter $%
\Gamma $ and doping. For the single crystal materials, one would identify
the main origin of disorder with the change of Coulomb potential at the Mn
sites due to substitution of trivalent rare earths by divalent alkaline
earths. This change amounts to some eV's, so that even including a
reasonable screening it remains within the order of magnitude of the
bandwidth. Crudely, we can assume that the different levels of doping
changes the center of the distribution of energies but do not affect $\Gamma
.$ In that case, samples of La$_{1-x}$Sr$_x$MnO$_3$ for example, differ only
in the hole concentration $x$. According to our results shown in Fig. 3, the
resistive behaviour of the samples with x=0.15, 0.175, 0.2, 0.3 as reported
in Ref. 16 by Tokura et. al. is consistent with a value of $\Gamma $ of about 2
(in units of the hopping parameter $t\sim 0.2eV$). Cation or Oxigen
vacancies should enhance the values of $\Gamma $ favoring e localization.
This is found in samples of (LaMn)$_{1-\gamma }$ O$_3$ as reported in Ref.
17 and in samples of La$_{1-x}$Ca$_x$MnO$_z$, Ref 18. To be able to compare
the temperature dependence or the resistivity with our results more
presicely it is necessary to obtain the magnetization and the energy
difference between the Fermi level and the mobility edge as functions of
temperature. We will publish these results elsewhere.

Other experimental results as transfer of optical weight ( that can be
infered from Fig. 2 ), or effect of pressure ( increase of $t$ ),
substitution of La by other rare earths ( decrease of $t$ ), also find a
simple and consistent explanation in terms of the model. As mentioned above,
polaronic effects are expected to be present in these materials as in other
perovskites in which electronic structure changes occour, but are not
necessary to the understanding of the phenomena under consideration.

In summary, we present here a simple model that can describe the properties
of La$_{1-x}$A$_x$MnO$_3$ and related compounds. It allows one to describe
transitions between ''metallic'' and ''insulating'' states by shifts of the
Fermi level relative to the mobility edge of the system as the magnetization
changes.

\vspace{7mm}

{\large {\bf FIGURE CAPTIONS }} \vspace{3mm}

Figure 1. Partial densities of states for spin up and down according to Eq.
(6) for $k=5$, $E_{\pm}= \mp 6$, $\bar{n}_{\sigma}^{\alpha} = 0$, $t=1$, and
different values of the magnetization.

Figure 2. Partial densities of states for spin up and down for the lower
band including a Lorentzian distribution of diagonal energies of width $%
\Gamma=1$, $k=5$, $t=1$, $n=0.88$, and $J>>t$. Bold lines indicate the zone
of energies where the states are extended. The Fermi level ($\epsilon_{F}$)
is indicated for each value of the magnetization.

Figure 3. Electron concentration versus magnetization "metal-insulator"
phase diagram for $k=5$, $t=1$ and different values of $\Gamma$.

Figure 4. Energy difference between paramagnetic and ferromagnetic phases in
units of $t$ for various values of $\Gamma$.


\begin{references}
\bibitem{1}  R. von Helmholt, J. Wecker, B. Holzapfel, L. Schultz, and K.
Samwer, Phys. Rev. Lett. {\bf 71}, 2331 (1993).

\bibitem{2}  G. H. Jonker and J. H. van Santen, Physica {\bf 16}, 337
(1950); J. H. van Santen and G. H. Jonker, Physica {\bf 16}, 599 (1950);

\bibitem{3}  Y. Moritomo, A. Asamitsu, and Y. Tokura, Phys. Rev. B {\bf 51},
16491 (1995); Y. Okimoto, T. Katsufuji, T. Ishikawa, A. Urushibara, T.
Arima, and Y. Tokura, Phys. Rev. Lett. {\bf 75}, 109 (1995); S. W. Cheong,
H. Y. Hwang, P. G. Radaelli, D. E. Cox, M. Marezio, B. Batlogg, P. Schiffer,
and A. P. Ramirez, Proceedings of the ''Physical Phenomena at High Magnetic
Fields - II'' Conference, Tallahassee, Florida. World Scientific, to be
published; M. Jaime, M. B. Salamon, K. Pettit, and M. Rubinstein, to be
published; M. C. Martin, G. Shirane, Y. Endoh, K. Hirota, Y. Moritomo, and
Y. Tokura, To be published; R. Mahendiran, R. Mahesh, A. K. Raichaudhuri,
and C. N. R. Rao, Solid State Commun. {\bf 94}, 515 (1995); H. L. Ju, J.
Gopalakrishnan, J. L. Peng, Qi Li, G. C. Xiong, T. Venkatesan, and R. L.
Greene, Phys. Rev. B {\bf 51}, 6143 (1995); Y. Tokura, A. Urushibara, Y.
Moritomo, T. Arima, A. Asamitsu, G. Kido, and N. Furukawa, J. Phys. Soc.
Jpn. {\bf 63}, 3931 (1994); M. K. Gubkin, A. V. Salesskii, V. G. Krivenko,
T. M. Perekalina, T. A. Khimich, and V. A. Chubarenko, JETP Lett. {\bf 60},
57 (1994).

\bibitem{4}  C. Zener, Phys. Rev. {\bf 82}, 403 (1951).

\bibitem{5}  P. W. Anderson and H. Hasegawa, Phys. Rev. {\bf 100}, 675
(1955).

\bibitem{6}  P. G. de Gennes, Phys. Rev. {\bf 118}, 141 (1960).

\bibitem{7}  K. Kubo and N. Ohata, J. Phys. Soc. Jpn. {\bf 33}, 21 (1972).

\bibitem{8}  J. Mazzaferro, C. A. Balseiro, and B. Alascio, J. Phys. Chem.
Solids {\bf 46}, 1339 (1985).

\bibitem{9}  N. Furukawa, J. Phys. Soc. Jpn. {\bf 63}, 3214 (1994).

\bibitem{10}  A. J. Millis, P. B. Littlewood, and B. I. Shrainman, SISSA:
condmat/9501034.

\bibitem{11}  T. Kasuya, Prog. Theor. Phys. {\bf 16}, 45 (1956). J. Phys.
Soc. Jpn, in press.

\bibitem{12}  See, e.g., E. N. Economou, {\it Green's Functions in Quantum
Physics} Springer Series in Solid-State Sciences {\bf 7}, Ed. P. Fulde.

\bibitem{13}  P. W. Anderson, Phys. Rev. {\bf 109}, 1492 (1958).

\bibitem{14}  D. C. Licciardello and E. N. Economou, Phys. Rev. {\bf 11},
3697 (1975).

\bibitem{15}  P. Lloyd, J. Phys. C {\bf 2}, 1717 (1969).


\bibitem{16} Y. Tokura, A. Urushibara, Y. Moritomo,
T. Arima, A. Asamitsu, G. Kido, and N. Furukawa, J. Phys. Soc.
Jpn. {\bf 63}, 3931 (1994).

\bibitem{17} L. Ranno, M. Viret, A. Mari, R. M. Thomas, and
J. M. D. Coey, J. Phys.: Condens. Matter {\bf 8}, L33 (1996).

\bibitem{18} H. L. Ju, J. Gopalakrishnan,
J. L. Peng, Qi Li, G. C. Xiong, T. Venkatesan, and R. L.
Greene, Phys. Rev. B {\bf 51}, 6143 (1995).
\end{references}
\end{document}